# Genesis of general relativity – Discovery of general relativity

Galina Weinstein

*The intermediate stage of the development of general relativity is inseparable of Marcel Grossmann's mathematical assistance. Einstein acknowledges Grossmann's help during 1912-1914 to the development of general relativity. In fact, as with special relativity so was it with General relativity, Einstein received assistance only from his old friends, Marcel Grossmann and Michele Besso. However, he continued to consider Besso as his eternal "sounding board"...*

In 1955 Einstein wrote his short *Autobiographical Sketch* dedicated mainly to his relations with his close friend Marcel Grossmann. In this *Skizze* Einstein told the story of his collaboration with Grossman which led to the Einstein-Grossmann theory. I will analyze Einstein's recollections, and while doing so outline Einstein's efforts at searching for a gravitational theory between 1912 and 1914.

**The line element**

Einstein had started to study Minkowski's four-dimensional reformulation of the special theory of relativity in earnest around 1910. However he did not use this formalism in his theory of static gravitational fields of 1912. Einstein wrote in his *Skizze*,[1]

"From the experience of the kind of scientific research which those happy Berne years have brought, I mention only one, the idea turned to be the most fruitful of my life. The special theory of relativity was several years old [...].

1909-1912, while I was teaching at the Zurich [1909-1910] and at the Prague universities of theoretical physics, I mused incessantly over the problem [gravitation]. 1912, when I was appointed to the Zurich Polytechnic, I was considerably closer to the solution of the problem. Of importance here proved to be Hermann Minkowski's analysis of the formal basis of the special theory of relativity".

Max Abraham told Einstein that in his theory of static gravitational fields of 1912 he failed to implement Minkowski's reformulation of special relativity (the "usual" theory of relativity as they put it) in terms of a four-dimensional space-time manifold. After the controversy with Abraham, Einstein realized that Minkowski's formalism was as crucial instrument for the further development of his theory of gravitation. However, a successful application of Minkowski's formalism to the problem of gravitation called for a mathematical generalization of this formalism.

---

[1] Einstein, Albert, "Erinnerungen-Souvenirs", *Schweizerische Hochschulzeitung* 28 (Sonderheft) (1955), pp. 145-148, pp. 151-153; Reprinted as, "Autobiographische Skizze" in Seelig Carl, *Helle Zeit – Dunkle Zeit. In memoriam Albert Einstein*, 1956, Zürich: Branschweig: Friedr. Vieweg Sohn/Europa, pp. 9-17; pp. 12-14.

In late spring 1912 Einstein found the appropriate starting point for such a generalization of Minkowski's formalism, the "line element" used in Minkowski's formalism which is invariant under the Lorentz group.

**The metric tensor**

Einstein went on to say in the *Skizze*, "[…] the Gravitational field is described by a metric, a symmetric tensor-field of metric $g_{ik}$".[2]

The crucial breakthrough had been that Einstein had recognized that the gravitational field should not be described by a variable speed of light as he had attempted to do in 1912 in Prague, but by the metric tensor field; a mathematical object of ten independent components, that characterizes the geometry of space and time.

Einstein wrote about his switch of attitude towards mathematics in the oft-quoted letter to Sommerfeld on October 29, 1912,[3]

"I am now occupied exclusively with the gravitational problem, and believe that I can overcome all difficulties with the help of a local mathematician friend. But one thing is certain, never before in my life have I troubled myself over anything so much, and that I have gained great respect for mathematics, whose more subtle parts I considered until now, in my ignorance, as pure luxury! Compared with this problem, the original theory of relativity is childish".

Einstein indeed was occupied with the more subtle parts of mathematics after adopting the metric tensor. He wrote in the *Skizze*,[4]

The problem of gravity was thus reduced to a purely mathematical one. Are there differential equations for *gik*, which are invariant under non-linear coordinate transformations? Such differential equations and *only* those were taken into consideration as the field equations of the gravitational field. The law of motion of material points was given by the equation of the geodesic line".

---

[2] Einstein, 1955, p. 15.
[3] Einstein to Sommerfeld, October 29, 1912, *The Collected Papers of Albert Einstein (CPAE) Vol. 5: The Swiss Years: Correspondence, 1902–1914*, Klein, Martin J., Kox, A.J., and Schulmann, Robert (eds.), Princeton: Princeton University Press, 1993, Doc. 421.
[4] Einstein, 1955, p. 15.

## The Einstein-Grossman collaboration

After arriving back to Zurich in summer 1912 Einstein was searching his "local mathematician friend" (as he told Sommerfeld above) from collage, Marcel Grossmann,[5]

"With this task in mind, in 1912, I was looking for my old student friend Marcel Grossmann, who had meanwhile become a professor of mathematics in the Swiss Federal Polytechnic institute. He was immediately caught in the fire, even though he had as a real mathematician a somewhat skeptical attitude towards physics. When we were both students, and we used to exchange thoughts in the Café, he once said such a beautiful and characteristic remark that I cannot help but quoting it here: 'I admit that from studying physics I have benefitted nothing essential. When I sat on the chair earlier and I felt a little of the heat that came from my "pre-seated" it grazed me a little. This has completely passed, because physics has taught me to consider writing that heat is something very impersonal'.

So he arrived and he was indeed happy to collaborate on the problem, but with the restriction that he would not be responsible for any statements and won't assume any interpretations of physical nature".

Einstein's collaboration with Marcel Grossmann led to two joint papers: the first of these was published before the end of June 1913,[6] and the second, almost a year later,[7] two months after Einstein's move to Berlin.[8]

Einstein and Grossmann's first joint paper entitled, "Entwurf einer verallgemeinerten Relativitätstheorie und einer Theorie der Gravitation" ("Outline of a Generalized Theory of Relativity and of a Theory of Gravitation") is called by scholars the "Entwurf" paper. Actually Einstein himself also called this paper "Entwurf", and he and Grossmann referred to the theory presented in this paper as the "Entwurf" theory:

---

[5] Einstein, 1955, pp. 15-16.
[6] Einstein, Albert, and Grossmann, Marcel, *Entwurf einer verallgemeinerten Relativitätstheorie und einer Theorie der Gravitation I. Physikalischer Teil von Albert Einstein. II. Mathematischer Teil von Marcel Grossman*, 1913, Leipzig and Berlin: B. G. Teubner. Reprinted with added "Bemerkungen", *Zeitschrift für Mathematik und Physik* 62, 1914, pp. 225-261. (*CPAE* 4, Doc. 13).
[7] Einstein, Albert and Grossmann, Marcel, "Kovarianzeigenschaften der Feldgleichungen der auf die verallgemeinerte Relativitätstheorie gegründeten Gravitationstheorie", *Zeitschrift für Mathematik und Physik* 63, 1914, pp. 215-225.
[8] *The Collected Papers of Albert Einstein. Vol. 4: The Swiss Years: Writings, 1912–1914*, Klein, Martin J., Kox, A.J., Renn, Jürgen, and Schulmann, Robert (eds.), Princeton: Princeton University Press, 1995, "Einstein on Gravitation and Relativity: The Collaboration with Marcel Grossman", p. 294.

in their second joint paper they wrote, "Gravitationsgleichung (21) bzw. (18) des 'Entwurfes'".[9]

Grossmann wrote the mathematical part of this paper and Einstein wrote the physical part. The paper was first published in 1913 by B. G. Teubner (Leipzig and Berlin). And then it was reprinted with added "Bemarkungen" (remark) in the *Zeitschrift für Mathematik und Physik* in 1914. The "Bemarkungen" was written by Einstein and contained the well-known "Hole Argument".[10]

The "Entwurf" theory was already very close to Einstein's general theory of relativity that he published in November 1915. The gravitational field is represented by a metric tensor, the mathematical apparatus of the theory is based on the work of Riemann, Christoffel, Ricci and Levi-Civita on differential covariants, and the action of gravity on other physical processes is represented by generally covariant equations (that is, in a form which remained unchanged under all coordinate transformations). However, there was a difference between the two theories, the "Entwurf" and general relativity. The "Entwurf" theory contained *different field equations* that represented the gravitational field, and these were *not* generally covariant. At first though – when Einstein first collaborated with Grossmann – he considered (in what scholars call the "Zurich Notebook) field equations that were very close to the ones he would eventually choose in November 1915.[11] But he gave them up in favor of the "Entwurf" field equations.

Starting in 1912, Einstein went through a long odyssey in the search after the correct form of the field equations of his new theory. The first trail of Einstein's efforts appear to be documented in a blue bound notebook – known as the "Zurich Notebook" – comprised of 96 pages, all written in Einstein's hand. The back cover of the Notebook bears the title "Relativität" in Einstein's hand (probably an indication that he began his notes at the end). Two pieces of paper were probably taped later to the front of the notebook by Einstein's secretary, Helen Dukas. The subject matter of the calculations in the notebook includes statistical physics, thermodynamics, the basic principles of the four-dimensional representation of electrodynamics, and the major part of the notebook is gravitation.

The notebook was found within Einstein's papers after his death. The calculations that Einstein had done in the final pages of the Notebook indicate continues path towards the "Entwurf" paper of 1913 with Marcel Grossman.

---

[9] Einstein and Grossmann, 1914, p. 217.
[10] Einstein and Grossman, 1913.
[11] *CPAE*, Vol.4, "Einstein on Gravitation and Relativity: The Collaboration with Marcel Grossmann", p. 294.

## Grossman brings to Einstein's attention the absolute Differential Calculus

At this very early stage during summer 1912 of calculations with the metric tensor, Einstein explained in the *Skizze* that Grossmann,[12]

"He looked through the literature, and soon discovered that the particular implied mathematical problem was already solved by Riemann, Ricci and Levi-Civita. The entire development followed the Gaussian theory of curvature-surfaces, which was the first systematical use of generalized coordinates. Reimann's achievement was the biggest. He showed how a field of *gik* tensors of the second differentiation rank can be formed".

About thirty years earlier, in 1922, Einstein was reported to have said, "I had the key idea of the analogy between the new mathematical problems connected with the new theory and the Gaussian theory of surfaces, however only after my return to Zurich in 1912, without at first studying Riemann and Ricci, as well as Levi-Civita. This was first brought to my attention by my friend Grossmann in Zurich, when I presented to him the problem of looking for generally covariant tensors whose components depend only on derivatives of the coefficients of the quadratic fundamental invariant [$g_{\mu\nu}$]".[13]

Einstein seemed to have had a key idea of the Gaussian theory of curvature-surfaces, as discussed below. Grossmann brought to his attention the works of Riemann, Ricci and Levi-Civita; and in addition he showed him the work of Christoffel. Louis Kollros, a professor of geometry and mathematics at the ETH, who was a fellow student of Einstein, wrote in memory of Einstein in 1955 that, sometime upon his arrival, Einstein spoke about his concern with Grossmann and told him one day, "Grossmann, you have to help me, or I shall go crazy! And Marcel Grossman had managed to show him that the mathematical tools he needed, were created just in Zurich in 1869 by Christoffel in the treatise 'On the Transformation of the Homogeneous differential Forms of the Second Degree', published in Volume 70. of the 'Journal de Crelle' for pure and applied mathematics".[14]

---

[12] Einstein, 1955, p. 16.
[13] Stachel, John, "Einstein's Search for General Covariance 1912-1915", in Howard, Don and Stachel John (eds), *Einstein and the History of General Relativity*, Einstein Studies, Vol 1, 1989, Birkhäuser, pp. 63-100; reprinted in Stachel, John, *Einstein from 'B' to 'Z'*, 2002, Washington D.C.: Birkhauser, pp. 301-338; p. 325.
[14] Kollros, Louis, "Erinnerungen eines Kommilitonen", 1955, in Seelig Carl, *Helle Zeit – Dunkle Zeit. In memoriam Albert Einstein*, 1956, Zürich: Branschweig: Friedr. Vieweg Sohn/Europa, p. 27.

**Geiser's lectures at the Polytechnic on Gauss' theory of curved surfaces**

Already before arriving at Zurich Einstein knew he needed a theory of invariants and covariants associated with the differential line element.[15] Kollros wrote, "Even in Prague he had foreseen that his generalized theory of relativity demanded much more than the mathematics of the elegant special relativity. He now found that elegance should rather remain a matter for tailors and shoemakers".[16]

Einstein was already aware of the need of the implementation of Minkowski's formalism and of Gauss' theory of curved surfaces. Marcel Grossmann's notebooks from the Polytechnic demonstrate that the lectures which Einstein heard as a student at school from Prof. Carl Friedrich Geiser had familiarized him with the Gaussian theory of two-dimensional surfaces.[17]

Einstein writes in his *Skizze*, "also I was fascinated by professor Geiser's lectures on infinitesimal Geometry, the true master pieces of art were pedagogical and helped very much afterwards in the struggle with the general theory of relativity".[18]

In the Kyoto lecture Einstein is reported to have said, "This problem was unsolved until 1912, when I hit upon the idea that the surface theory of Karl Friedrich Gauss might be the key to this mystery. I found that Gauss' surface coordinates were very meaningful for understanding this problem. Until then I did not know that Bernhard Riemann had discussed the foundation of geometry deeply. I happened to remember the lecture on geometry in my student years [in Zurich] by Carl Friedrich Geiser who discussed the Gauss theory. I found that the foundations of geometry had deep physical meaning in this problem.

When I came back to Zurich from Prague, my friend the mathematician Marcel Grossmann was waiting for me. […] First he taught me the work of Curbastro Gregorio Ricci and later the work of Riemann".[19]

**The Riemann Tensor and the Ricci Tensor**

Pais recalls a discussion with Einstein in which he asked Einstein how the collaboration with Grossmann began."I have a vivid though not verbatim memory of Einstein's reply: he told Grossman of his problems and asked him to please go to the library and see if there existed an appropriate geometry to handle such questioning.

---

[15] Pais, Abraham, *Subtle is the Lord. The Science and Life of Albert Einstein*, 1982, Oxford: Oxford University Press, p. 212.
[16] Kollros, 1955, p. 27.
[17] Stachel, 1989, in Stachel 2002, p. 303.
[18] Einstein, 1955, pp. 10-11.
[19] Einstein, Albert, "How I Created the Theory of Relativity, translation to English by. Yoshimasha A. Ono, *Physics Today* 35, 1982, pp. 45-47, p. 47.

The next day Grossman returned (Einstein told me) and said that there indeed was such a geometry, Riemannian geometry".[20]

Grossmann helped Einstein in his search for a gravitational tensor, and like Grossmann Einstein "was immediately caught in the fire". Just before writing the "Entwurf" paper with Grossmann, Einstein struggled with these new tools in a small blue Notebook –named by scholars the "Zurich Notebook".[21] Einstein filled 43 pages of this notebook with calculations, while he was fascinated with Riemann's calculus.

While filling the notebook he received from time to time the new mathematical tools from Grossmann, and he wrote Grossmann's name in the notebook every time he got something new to indicate the tensors that he received from him. At the top of the 14L page Einstein wrote on the left: "Grossmann's tensor four-manifold" and next to it on the right he wrote the fully covariant form of the Riemann tensor. On top of the 22R page he wrote Grossmann's name. [22]

On page 14L of the Zurich Notebook Einstein systematically started to explore the Riemann tensor. It appears that Grossman suggested the four-rank Riemann tensor as a starting point. In the course of this exploration Einstein considered on page 22R candidate field equations with a gravitational tensor that is constructed from the Ricci tensor; an equation Einstein would come back to in November 1915.

in the Zurich Notebook Einstein searched for a gravitational theory, and a gravitational field equation, that would satisfy some heuristic requirements. Einstein had to cope with new mathematical tools, and at the same time he was guided by few heuristic principles: the principle of relativity, the equivalence principle, the correspondence principle, and the principles of conservation of energy and momentum.[23]

In the Zurich Notebook Einstein first tackled relativity and equivalence and arrived at general covariance, and then moved on to correspondence and conservation. But afterwards it was just the other way round. Einstein first tackled correspondence and conservation and then relativity and equivalence, and lost general covariance. The interplay of the four heuristic principles with the new tools of absolute differential calculus of 1912 that Einstein was exploring governed the form of the field equations Einstein was finally left with at the end of the Zurich Notebook. Hence, at the end of the day Einstein's conditions over determined Einstein's research between 1912 and 1913.

---

[20] Pais, 1982, p. 212.
[21] *CPAE*, vol. 4, Doc. 10.
[22] *CPAE*, Vol.4, "Einstein on Gravitation and Relativity: The Collaboration with Marcel Grossmann", p. 296.
[23] Renn, Jürgen and Sauer, Tilman, "Pathways out of Classical Physics. Einstein's Double Strategy in his Search for the Gravitational Field Equation", in Janssen, Michel, Norton, John, Renn, Jürgen, Sauer, Tilman, Stachel John, *The Genesis of General Relativity* ed. Renn, Jürgen, *Vol. 1: Einstein's Zurich Notebook, Introduction and Source*, 2007, New York, Berlin: Springer, pp. 113-312; pp. 123-125; Renn, Jürgen and Tilman Sauer, "Heuristics and mathematical Representation in Einstein Search for a Gravitational Field Equation", Preprint 62, *Max Planck Institute for the History of Science*, 1997, p. 13-17; Janssen, Michel, Norton, John, Renn, Jürgen, Sauer, Tilman, Stachel John, ed., *The Genesis of General Relativity*. 4 Vols., 2007, New York, Berlin: Springer, p. 500.

Renn and Sauer formulated Einstein's heuristic principles that played a role in Einstein's search and rejection of generally covariant field equations between 1912 and 1913. They explain that "Each of Einstein's heuristic principles against which constructions would have to be checked could be used either as a construction principle or as a criterion for their validity":[24]

1) *The Equivalence principle*: After 1912 Einstein implemented the equivalence principle by letting the metric field $g_{\mu\nu}$ represent both the gravitational field and the inertial structure of space-time. The metric field is a solution of the same field equations in the coordinate systems of both observers. This is automatically true if the field equations are generally covariant. But if the field equations are of restricted covariance, then Einstein tried to implement the equivalence principle in other ways. The equivalence principle became the principle core of Einstein's search for a generalization of the relativity principle for non-uniform motion.[25]

2) *The generalized principle of relativity*: From 1912 onward, Einstein attempted to generalize the principle of relativity by requiring that the covariance group of his new theory of gravitation be larger than the group of Lorentz transformations of special relativity. In his understanding this requirement was optimally satisfied if the field equation of the new theory could be shown to possess the mathematical property of general covariance.[26]

Einstein was thus under the impression that the principle of relativity for uniform motion of special relativity could be generalized to arbitrary motion if the field equations possessed the mathematical property of general covariance. And if the principle of relativity is generalized then the equivalence principle is satisfied, because according to this principle, an arbitrary accelerated reference frame in Minkowski space-time can precisely be considered as being physically equivalent to an inertial reference frame if a gravitational field can be introduced which accounts for the inertial effects in the accelerated frame. Einstein then tried to construct field equations of the broadest possible covariance.[27] In the Zurich Notebook Einstein succeeded in formulating a generally-covariant equation of motion for a test particle in an arbitrary field. In this equation, the gravitational potential is represented by a four-dimensional metric tensor, which became the key object for Einstein's further research in the following years.[28]

---

[24] Renn and Sauer, 2007, in Renn (2007), Vol. 1, p. 151.
[25] Renn and Sauer, 1997, p. 15.
[26] Renn and Sauer, 1997, p. 15.
[27] Janssen, Norton, Renn, Sauer, and Stachel 2007, in Renn (2007), p. 494.

3) *The conservation principle*: The energy-momentum conservation principle played a crucial role in Einstein's static gravitational theory of 1912. Einstein started with the relation between mass and energy from special relativity. He extended it within his theory of static gravitational fields. However, scientists were already extending special relativity to relativistic dynamics, and conservation of energy and momentum centered at that time upon a four-dimensional stress-energy tensor. Starting in 1912 Einstein embodied the mass and energy relation in the energy-momentum tensor as the source of the gravitational field. Einstein also required that the gravitational field equation should be compatible with the generalized requirement of energy and momentum conservation.[29] In 1912 this caused Einstein to think that energy-momentum conservation required that the covariance of the field equations be restricted.[30]

4) *The correspondence principle*: Einstein was searching in two directions: he was looking for an equation of motion for bodies in a gravitational field and for a field equation determining the gravitational field itself, the generalization of the Poisson equation.[31] The Poisson equation of classical gravitation theory describes how gravitating matter generates a gravitational potential. This potential can then be related to the gravitational field and to the force acting on the material particles exposed to it. Einstein's theory of static gravitational fields provided a starting point, since it represents a step beyond the Poisson equation towards a generalized relativistic theory. One of Einstein's earliest attempts in his Zurich notebook was to construct equations that would mimic the way in which the classical Laplace operator was formed.[32]

However, Einstein was examining various candidates for generally covariant field equations in his Zurich Notebook; and he wanted to check whether these unknown gravitational field equations for the metric tensor fulfilled the requirement of recovering the familiar Newtonian gravitation theory (the Poisson equation of Newtonian theory for the scalar Newtonian potential) in the special case of the low velocity limit and weak static gravitational field. Einstein expected that under appropriate limiting conditions, the theory he was developing would first reproduce the results of his 1912 theory of static gravitational fields; that is, he expected his new theory would reduce to his own earlier theory of static fields in which gravitational potential is represented by a variable speed of light. And then, under further constraints, he would be able to recover the classical Poisson equation.[33]

---

[28] Renn and Sauer, 2007, in Renn (2007), vol. 1, p. 123.
[29] Renn and Sauer, 2007, in Renn (2007), vol. 1, p. 147.
[30] Janssen and Renn, 2007 in Renn (2007), vol. 2, p. 842.
[31] Renn and Sauer, 2007, in Renn (2007), vol. 1, p. 129.
[32] Renn and Sauer, 2007, in Renn (2007), vol. 1, p. 162.

While searching for the gravitational field equation, Einstein had an additional problem, to ensure the compatibility of the different heuristic requirements by integrating them into a coherent gravitation theory represented by a consistent tensorial framework.[34] This complexity led to the following situation: The correspondence principle became the weightiest of Einstein's heuristic principles.[35]

At this stage Einstein explored a different candidate for the metric tensor, and implemented differently the correspondence principle. And in fact he eventually perused all options. Of course he could not give up so easily the conservation principle and the settings of the stress-energy tensor. He had to solve the conflict between them in order to solve the incompatibility problem between the correspondence and conservation principles. Einstein changed the left hand side of the field equations and solved the conflict between the two above principles, but had to check whether the conservation principle was fully satisfied. Once this issue was settled the correspondence principle caused problems with the theory of static gravitational fields that was obtained using the equivalence principle.[36]

At the end the match between the correspondence principle and the conservation principle was achieved at the expense of the generalized principle of relativity.[37] At some stage, thus, Einstein appeared to somewhat forgot a little from the generalized principle of relativity; the starting point of his research project. Meanwhile he had developed many tools that would finally lead him to the goal of generally covariant equations. But at this stage he was not quite sure about the most important principles his novel theory should fulfill. He found it difficult to establish a match between the equations and the principles. But during 1912-1913 he gradually created the tensorial framework for his future general relativity.

It appears that Einstein's long vacillating between general covariance and the correspondence principle is a symptom of his fixation on the older 1912 paradigm of the static gravitational fields. He needed an extra two years to gradually and mentally switch into the new paradigm of differential calculus; yet he would always envision Riemann's calculus in terms of his heuristic principles.

The Zurich Notebook shows that Einstein already considered the field equations of general relativity about three years before he published them in November 1915.[38] Einstein had come close to his November 4, 1915 field equation on page 22R when he considered the Ricci tensor given to him by Grossmann as a possible candidate for the left hand-side of such a field equation.[39] As we have seen, the analysis of the notebook by the scholars even revealed that Einstein got so close to his November 4, 1915 breakthrough at the end of 1912, that he even considered on page 20L another candidate – albeit in a linearized form – which resembles the final version of the November 25, 1915 field equation of general relativity $[R_{\mu\nu} - (1/2)g_{\mu\nu}R]$.[40]

---

[33] Renn and Sauer, 2007, in Renn (2007), vol. 1, p. 148.
[34] Renn and Sauer, 2007, in Renn (2007), vol. 1, p. 125.
[35] Renn and Sauer, 2007, in Renn (2007), vol. 1, p. 149.
[36] Renn and Sauer, 2007, in Renn (2007), vol. 1, pp. 217-218.
[37] Renn and Sauer, 2007, in Renn (2007), vol. 1, p. 225.
[38] Janssen and Renn, 2007 in Renn (2007), vol. 2, p. 839.

Thus Einstein first wrote down a mathematical expression close to the correct field equation and then discarded it, only to return to it more than three years later. Scholars asked: Why did Einstein discard in winter 1912-1913 what appears in hindsight to be essentially the correct gravitational field equation, and what made his field equation acceptable in late 1915?[41] Why did he reject equations of much broader covariance in 1912-1913?[42] The answer the scholars gave was that accepting the correct mathematical expression in 1912 required abandoning a few heuristic conditions that Einstein could not yet reconcile with his field equations of 1912. This path, says Norton, was reasonable at the time, 1912-1913. His rejection of the Ricci tensor need not be explained in terms of simple error. He was rather not prepared to accept generally covariant equations as a result of a number of misconceptions.[43]

Einstein's work in the Zurich Notebook was so laborious and painful but he did think that this work was rewarded when he published the "Entwurf" paper. The scholars jointly summarized the situation in the spring of 1913 in the following way: Through trial and error Einstein found a body of results, strategies, and techniques that he drew on for the more systematic search for field equations. He checked a series of candidate field equations against a list of criteria they would have to satisfy. All these candidates were extracted from the Riemann tensor. In the Zurich Notebook Einstein eventually gave up trying to extract field equations from the Riemann tensor. Drawing on results and techniques found during the earlier stage instead, he developed a way of generating field equations guaranteed to meet what he deemed to be the most important of the requirements to be satisfied of such equations. In this way he found the field equations of severely limited covariance published in the spring of 1913 in the paper co-authored with Marcel Grossmann, the "*Entwurf*" paper.

**The "Entwurf" field equations**

On May 1913 Einstein was more confident that he was probably not deceived with the "Entwurf" theory of him and Grossman. Einstein is reported to have said in the Kyoto lecture "I discussed with him [Grossmann] whether the problem could be solved using

---

[39] Renn (2007), Vol. 2, pp. 451; 453.
[40] Janssen, Norton, Renn, Sauer, and Stachel, 2007, in Renn (2007), p. 634.
[41] Renn and Sauer, 2007, in Renn (2007), vol. 1, p. 118.
[42] Janssen and Renn, 2007 in Renn (2007), vol. 2, p. 839.
[43] Norton, John, "How Einstein Found His Field Equations: 1912-1915," *Historical Studies in the Physical Sciences* 14, 1984, pp. 253-315; p. 255.

Riemann theory, in other words, by using the concept of invariance of line elements. We wrote a paper on this subject in 1913, although we could not obtain the correct equations for gravity. I studied Riemann's equations further only to find many reasons why the desired results could not be attained in this way".[44]

At first, Einstein was not quite satisfied with the "Entwurf" equations. He wrote his friend Paul Ehrenfest in May 1913,

"I am now deeply convinced of having taken the thing right, and also, of course, that a murmur of indignation will spread through the ranks of our colleagues when the work ["Entwurf"] appears, which will take place in a few weeks. Naturally, I will send you right away a copy. The conviction which I have slowly struggled my way through is that *privileged coordinate systems do not exist at all*. However, I succeeded only partially, to formally penetrate to this position as well".[45]

By this time Einstein knew he only succeeded partially from the formal point of view – he did not possess generally covariant field equations. However, to use Einstein's own chair story that he told of Grossmann before in the *Skizze*, now when Einstein was going to sit on his new honored chair, and feel a little of the heat that came from his "pre-seated" giant, Newton, and this heat grazed him a little; it did not completely pass. Because physics has taught him to consider writing that the Newtonian limit was something very important.

At the beginning Einstein was confused from the abrupt transition from scalar theory to tensors: "I thought again about the scalar theory [from Prague] when I was at first a bit overawed by the complexity of the equations which Grossmann and I wrote down a little later. Yes there was confusion at that time, too. But it was not like the Prague days. In Zurich I was sure that I had found the right starting point".[46]

Conservation of momentum and energy is also a basic pillar of physics. This brought problems when creating a theory in which there were generally covariant equations that completely determined the gravitational field from the matter tensor. Einstein was at first dissatisfied with the lack of general covariance of his field equations. He did not have much faith in the new theory. He continued to search for the most general transformation under which the gravitational field equations of the "Entwurf" theory are invariant, but by mid August he was still unable to find a single non-linear transformation admitted by the "Entwurf" field equations.[47]

---

[44] Einstein, 1922/1982, p. 47.
[45] Einstein to Ehrenfest, May 28, 1913, *CPAE*, vol. 5., Doc. 441.
[46] Pais, 1982, p. 204.
[47] *CPAE*, Vol.4, "Einstein on Gravitation and Relativity: The Collaboration with Marcel Grossmann", p. 297; Einstein to Lorentz, 14 August, 1913, *CPAE*, Vol. 5, Doc. 467.

**The Hole Argument**

Einstein thought for a while – or persuaded himself – that generally covariant field equations were not permissible; one must restrict the covariance of the equations. He gave two arguments for this:

1) Einstein supposed that the stress-energy tensors for matter and for gravitation are generally covariant. He then considered the conservation law for matter and gravitational field together. However, Einstein realized that this expression cannot be covariant with respect to arbitrary transformations. Therefore, by formulating this expression Einstein created an equation which was restricted. He restricted himself to privileged systems in which the law of conservation of momentum and energy holds in the form he had written it. If one privileges a group of systems, it means that the conservation law for matter and gravitational field is only covariant with respect to arbitrary linear transformations.[48]

2) He introduced an ingenious argument – the Hole Argument – to demonstrate that generally covariant field equations were not permissible. The Hole Argument seemed to cause Einstein great satisfaction, or else he persuaded himself that he was satisfied. Having found the Hole argument, Einstein spent two years after 1913 looking for a non-generally covariant formulation of gravitational field equations.[49]

Around this time, Einstein used to write to his friends, I am completely satisfied now from my theory of gravitation, and I no longer have doubts of the correctness of the theory.[50] Although he thought he had invoked an ingenious demonstration, according to which it was unavoidable that the gravitational equations were not generally covariant with respect to arbitrary transformations, the issue seemed to still bother him very much. He wrote Hopf on November 1913 that this topic bothered him so much for so long.[51] But now it was settled. Einstein needed the support of his friends when struggling with the development of the new theory of gravitation. He was rather isolated because physicists (and the highest in ranks, Plank) behaved passively towards his new gravitation theory.[52] He felt he was on the right track with the equivalence principle, and the equality of inertial and gravitational mass, and this caused him great satisfaction. But he did have this inkling that something was missing. He wrote Zangger on March 10, 1914,

---

[48] Einstein to Lorentz, August 16, 1913, *CPAE*, Vol. 5, Doc. 470.
[49] Satchel, "The Rise and Fall of Einstein's Hole Argument".
[50] Einstein to Ehrenfest, November 7, 1913, *CPAE*, Vol. 5, Doc. 481; Einstein to Ehrenfest, second half of November, 1913, *CPAE*, Vol. 5, Doc. 484; Einstein to Zangger, March 10, 1914, *CPAE*, Vol. 5, Doc. 513; Einstein to Besso, March 10, 1914, *CPAE*, Vol. 5, Doc. 514.
[51] Einstein to Hopf, November 2, 1913, *CPAE*, Vol. 5, Doc. 480.
[52] Einstein to Besso, January 1, 1914, *CPAE*, Vol. 5, Doc. 499.

"Now the harmony of the mutual relationship in the theory is such that I no longer have the slightest doubt about its correctness. Nature shows us only the tail of the lion. But I have no doubt that the lion belongs with it even if he cannot reveal himself to the eye directly, because of his huge dimension".[53] It appears that Einstein was aware that at this point he could only see a little of the tail of the lion; but he did not have the slightest doubt of the correctness of the main scaffold of his physical theory and of the direction that he chose: the absolute differential geometry.

**Mercury, Rotation, and Light Deflection**

Einstein's collaboration with his close friends included Michele Besso as well. During a visit by Besso to Einstein in Zurich in June 1913 they both tried to solve the "Entwurf" field equations to find the perihelion advance of Mercury in the field of a static sun in what is known by the name, the "Einstein-Besso manuscript".[54] Besso was inducted by Einstein into the necessary calculations. The "Entwurf" theory predicted a perihelion advance of about 18" per century instead of 43" per century.[55] As shown later, Einstein did not mention Besso's name in his 1915 paper that explains the anomalous precession of Mercury. Besso collaborated with Einstein on the wrong gravitational (the "Enrwurf") theory, and their calculation based on this theory gave a wrong result. Towards the end of 1915 Einstein abandoned the "Entwurf" theory, and with his new theory got the correct precession so quickly because he was able to apply the methods he had already worked out two years earlier with Besso. Einstein though did not acknowledge his earlier work with Besso.

It appears that Einstein did not mention Besso because he considered him as a sounding board even though Besso was calculating with Einstein in Zurich; this as opposed to his other friend, Marcel Grossmann, who was his active partner since 1912 in creating the "Entwurf" theory. Einstein wrote Besso a series of letters between 1913 and 1916, and described to him step by step his discoveries of General Relativity, and thus Besso functioned again as the good old sounding board as before 1905.[56]

Alberto Martínez objected to characterizing Besso as a "sounding board" for Einstein's ideas. Martínez wrote, that this description was "first used by Einstein but repudiated by Besso as downplaying his role in their discussions and collaborations".[57] However, it seems that from 1912 Einstein considered Grossmann,

---

[53] Einstein to Zangger, March 10, 1914, *CPAE*, Vol. 5, Doc. 513.
[54] *CPAE*, Vol. 4, Doc 14.
[55] *CPAE*, Vol. 4, "The Einstein-Besso Manuscript on the Motion of the Perihelion of Mercury", p. 351.
[56] Einstein, Albert and Besso, Michele, *Correspondence 1903-1955* translated by Pierre Speziali, 1971, Paris: Hermann, Letters 9 to 14.
[57] Martínez, Alberto, "Review of John Stachel's *Einstein from 'B' to 'Z'*", *Physics in Perspective* 5, 2003, pp. 352-354, p. 354.

and not Besso, as his partner; and Besso remained Einstein's closest friend and sounding board.

On pages 41-42 of the Einstein-Besso manuscript, Einstein checked whether the metric field describing space and time for a rotating system was a solution of the field equations of the "Entwurf" theory. Einstein's answer was yes, but he later discovered that he made a mistake in the calculations.

Like its predecessor for the static gravitational field from 1911-1912, the "Entwurf" theory predicted the same value for the deflection of light in a gravitational field of the sun, 0.83 seconds of arc. Einstein, however, could not yet send an expedition to check this prediction of his theory. He made many efforts to obtain empirical data on light deflection, first from already existing photographs and later by involving himself in the organization of an expedition for the 1914 total solar eclipse; but the empirical verification of the light-bending effect remained elusive until 1919.[58]

Meanwhile Gunnar Nordström accepted the equivalence of inertial and gravitational masses. His theory of gravitation was more natural than the "Entwurf" theory, simpler, and more related to the original theory of relativity, and to its light postulate. According to Nordström there existed a red shift of spectral lines, as in Einstein's "Entwurf" theory, but there was no bending of light rays in a gravitational field. It could also not explain the anomalous motion of Mercury. But Nordström's theory became a true option for a gravitational theory.

At that time the "Entwurf" theory remained without empirical support. Thus a decision in favor of one or the other theory – the "Entwurf" or Nordström's – was impossible on empirical grounds.[59] Einstein began to study Nordström's theory from the theoretical point of view.[60] He realized that it did not satisfy Mach's ideas: according to this theory, the inertia of bodies seems to not be caused by other bodies. In a joint 1914 paper with Lorentz's student Adrian Fokker, Einstein showed that a generally covariant formalism is presented from which Nordström's theory follows if a single assumption is made that it is possible to choose preferred systems of reference in such a way that the velocity of light is constant; and this was done after Einstein had failed to develop a generally covariant formulation for the "Entwurf" theory.[61]

---

[58] *CPAE*, "Einstein on Grossmann and Relativity: The Collaboration with Marcel Grossmann", Vol. 4, p. 295.
[59] *CPAE*, "Einstein on Grossmann and Relativity: The Collaboration with Marcel Grossmann", Vol. 4, pp. 295; 299.
[60] Einstein, Albert, "Zum gegenwärtigen Stande des Gravitationsproblems", *Physikalische Zeitschrift* 14, 1913, pp. 1249-1262; Einstein, Albert and Fokker, Adriann, D., "Die Nordströmsche Gravitationstheorie vom Standpunkt des absoluten Differentialkalküls", *Annelen der Physik* 44, 1914, pp. 321-328.
[61] *CPAE*, "Einstein on Grossmann and Relativity: The Collaboration with Marcel Grossmann", Vol. 4, pp. 299-300.

**Einstein and Grossmann's second paper**

Meanwhile Einstein and Grossmann wrote their second paper, published almost a year after the "Entwurf" paper. They elaborated the gravitational "Entwurf" field equations. At this stage Einstein possessed the Hole Argument, and Einstein convinced himself by calculations that the "Entwurf" equations hold good in a uniformly rotating system. Einstein's desire was that acceleration-transformations – nonlinear transformations – would become permissible transformations in his theory. In this way transformations to accelerated frames of reference would be allowed and the theory could generalize the principle of relativity for uniform motions. Einstein wrote Besso before publishing the paper,

 "This shows that there exists acceleration transformations of varied kinds, which transform the equations to themselves (e.g. also rotation), so that the equivalence hypothesis is preserved in its original form, even to an unexpectedly large extent". Einstein was so happy that he concluded, "Now I am perfectly satisfied and no longer doubt the correctness of the whole system, regardless of whether the observation of the solar eclipse will be successful or not. The logic of the thing is too evident".[62] But Einstein continued on the one hand to struggle with the problem of generally covariant field equations, and on the other, with arguments (like the Hole Argument) intended to explain why he could not arrive at such equations.

**End of collaboration with Grossmann**

Einstein left Zurich in March-April 1914, and by this ended his collaboration with Marcel Grossmann. After real hard work in Berlin Einstein would finally find a brilliant way to create a theory, in which the generally covariant field equations he had already obtained in 1912 (in the Zurich Notebook) are interwoven with his physics of principles.

Einstein wrote towards the end of the *Skizze*,

"While I was busy at work together with my old friend, none of us thought of that tricky suffering, now this noble man is deceased. The courage to write this little colorful Autobiographical Sketch, gave me the desire to express at least once in life my gratitude to Marcel Grossmann".[63]

---

[62] Einstein to Besso, March 10, 1914, *CPAE*, Vol. 5, Doc. 514.
[63] Einstein, 1955, p. 16.